\documentclass[a4paper]{article}
\usepackage{a4wide}
\usepackage{physics}
\usepackage{graphicx}

\def\gg    {$\gamma \gamma$} 
\def\gge   {$\gamma ^* \gamma ^*$} 
\def\ggh    {$\gamma \gamma \rightarrow$ {\sl hadrons}}
\def\sggh    {$\sigma(\gamma \gamma \rightarrow$ {\sl hadrons})}

\def\sgg    {$\sigma _{\gamma \gamma }$}
\def\csee {$ \sigma ($ \ee \ra {} \ee {\sl hadrons})} 
\def\seeh {$d \sigma ($ \ee \ra {} \ee {\sl hadrons})}

\def\eett {  \ee \ra {}  \ee $\tau \tau $}
\def\ggrr {  \gg \ra {}   $\rho ^0 \rho ^0$}
\def\eepipi {  \ee \ra  {} \ee $\pi ^+ \pi ^- \pi ^+ \pi ^-$}

\def\Lgg  {$\cal{L}_{\gamma \gamma }$}

\def\Wgg {$W_{\gamma \gamma }$}
\def\Wvis {$W_{\mathrm{vis}}$}

\def\Q2 {$Q^2$}
\def\q2 {$q^2$}
\newcommand{\AmS}{{\protect\the\textfont2
  A\kern-.1667em\lower.5ex\hbox{M}\kern-.125emS}}

\title{Hadron production in two-photon collisions at
LEP-L3\footnote{Presented at the Central European Triangle Symposium on
Particle Physics, Zagreb, Croatia, June 17-19, 1999}
}

\author{\'Akos Csilling\\
KFKI Research Institute for Particle and
Nuclear Physics \\ 
 H-1525 Budapest, P.O.Box 49, Hungary
        }

\date{\null}       

\begin{document}

% typeset front matter (including abstract)
\maketitle

\begin{abstract}

The reaction 
 $\mathrm{e}^{+} \mathrm{e}^{-} \rightarrow \mathrm{e}^{+} \mathrm{e}^{-}
 \gamma ^{*}  \gamma ^{*}  \rightarrow 
 \mathrm{e}^{+} \mathrm{e}^{-} $ {\sl hadrons}
is analysed for quasi-real virtual photons using 
data collected by the L3 detector  during the LEP high energy runs 
at $ \sqrt {s} =$ 183 and   189 GeV.
 Preliminary results on the cross sections 
 $\sigma(\mathrm{e}^{+} \mathrm{e}^{-} \rightarrow 
 \mathrm{e}^{+} \mathrm{e}^{-} $ {\sl hadrons}) and  
$\sigma (\gamma\gamma \rightarrow $ {\sl hadrons}) are given in the 
interval 5 GeV $\leq W_{\gamma\gamma} \leq$ 145 GeV.
 The centre-of-mass energy dependence of the two-photon
 cross section 
is well described by the universal Regge parametrisation,
but with a steeper rise with energy as compared to
hadron-hadron cross sections. The data are 
also compared to the expectations of different theoretical models.
To investigate diffractive processes,
the elastic $\gamma\gamma\to\rho^0\rho^0$
process and  the inclusive $\rho^0$
production $\gamma\gamma\to\rho^0X$ are studied in  
the $W_{\gamma\gamma}\ge3$ GeV region.
In all channels  a comparison is made with
the {\sc Pythia} and {\sc Phojet} Monte Carlo generators.  
\end{abstract}

\vspace*{-11cm}

\null\hfill
KFKI-1999-08/A\\
\null\hfill\vspace{10cm}
October 25, 1999

\section{INTRODUCTION}

The \ee \ra \ee \gge \ra \ee {\sl hadrons} process is  a copious
source of hadron production at high energies. The hadron system has
predominantly a low mass value.  A large fraction of the hadrons
escape detection due to the large diffractive cross section and the
Lorentz boost of the \gg {} system. For these events, the measured
effective mass \Wvis{} is  smaller than the centre of mass energy of
the two interacting photons \Wgg. In this reaction  most of the
initial energy is taken by the scattered  electrons and positrons. 
We analyse  only data where these are not detected (anti-tagged
events). 

New results are presented on the total cross section \csee {} for
the  \ee {} centre of mass energy of   \rts = 183 \GeV {} and \rts =
189 \GeV. The  two-photon cross section \sggh {} is then derived in
the   interval 5 $\leq W_{\gamma \gamma} \leq$ 145 \GeV. The
analysis of the data taken at \rts = 130-161 \GeV{} has been
published~\cite{had161}.

\section{TOTAL CROSS-SECTION}

\subsection{Event selection}
\label{sel}

Data have been  collected with the L3 detector~\cite{L3,MSD} at \rts =
182.72 \GeV {} with a total integrated luminosity of 51.35 \pb \ during
1997 and at \rts = 189 \GeV {}  with a total integrated luminosity of
171.8 \pb \ during 1998.

Hadronic two-photon
  events are selected by the following criteria :

\begin{list}{$\bullet$}{
\setlength{\topsep}{0pt}
\setlength{\itemsep}{0pt}
\setlength{\parsep}{0pt}
}  
\item 
At  least six particles must be detected.
A particle can be  a track, a photon or a pion
cluster in the hadron calorimeter or 
in the luminosity monitor. 

\item  The total energy in the electro-magnetic calori\-meter is
required to be  greater than 500 \MeV. The total  energy deposited in
the  calorimeters must be  below 40\% of the collision
energy. 

\item
 An  anti-tag condition excludes events
  with an electromagnetic 
  shower shape and energy greater than 70 \GeV
{} in the luminosity monitor, 
in a fiducial region of
 33-64 mrad. The $Q^2$ value of
 the selected events is thus smaller than 8 $\GeV ^2$, in average
  $\langle Q^2\rangle \sim 1.5$ $ 10^{-2} \GeV ^2$.

\end{list}
After  selection, the background from beam-gas and beam-wall
interactions is found to be negligible. 
 The analysis is limited to events with \Wvis $\geq$ 5 \GeV.
More than 1 million events are selected. 
% 159012 at 183, 888170 at 189

\begin{figure}
\centering
  \includegraphics[width=0.75\textwidth,height=0.7\textwidth]{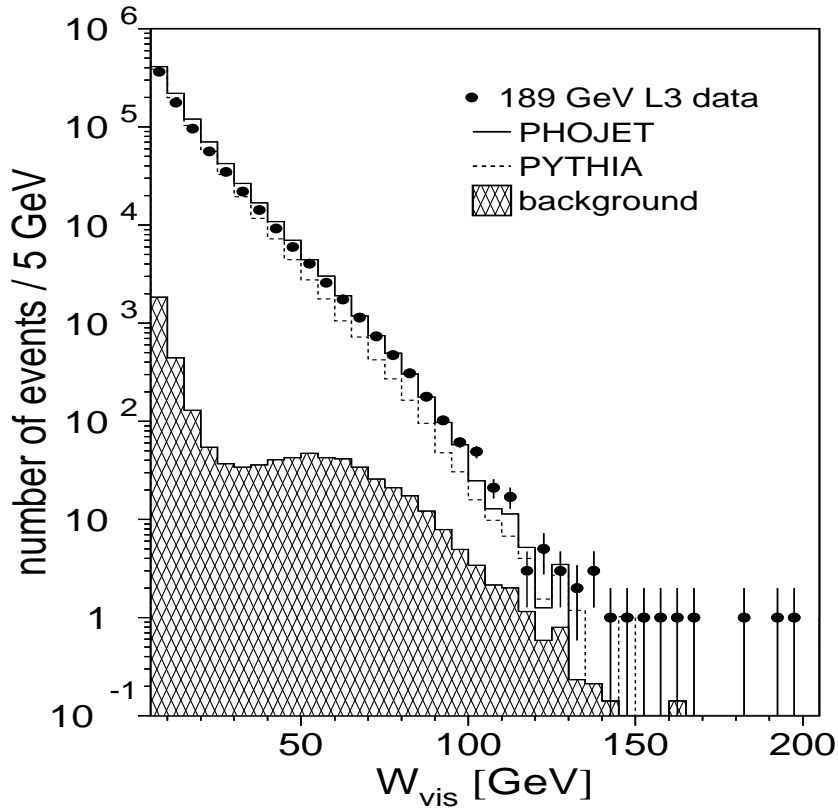}
 \caption{The measured hadronic mass \Wvis. 
 The backgrounds due to \ee annihilation and \eett {} 
  are indicated as a shaded area.} 
  \label{unfa}
\end{figure}

The various \ggh {} processes are simulated with the  {\sc
Phojet}\footnote{{\sc Phojet} version1.05c} \cite{Engel} and {\sc
Pythia}\footnote{{\sc Pythia} version 5.718 and {\sc Jetset} version 7.408}
\cite{pythia}  event generators. The background processes \ee
$\rightarrow$ {\sl hadrons}($\gamma $), ZZ($\gamma $), Zee($\gamma $),
We$ \nu$($\gamma $), $\tau^{+} \tau^{-}(\gamma )$, $W^{+} W^{-}$ and
\ee $\tau^{+} \tau^{-}$ were simulated and subtracted. At low masses
the background is below 1\%, dominated by  the two-photon $\tau$
production. It increases at  high masses, due  to annihilation
processes, and reaches  a maximum of 12\%  at 145 \GeV. In
Figure~\ref{unfa}  the \Wvis {} spectrum is shown for the \rts = 189
\GeV\ data. The trigger efficiency is mass  dependent, for  the \rts =
189 \GeV {} data it varies from 78\% at \Wvis =5 \GeV {} to 94\% for
\Wvis $ >$ 80 \GeV.

\subsection{Unfolding and efficiency}
From the observed distribution of the visible
effective mass, \Wvis,  the true hadron mass \Wgg {}
distribution must be extracted. The number of  observed
events are then 
corrected for the efficiency and acceptance of the detector.

 The measured \Wvis {} spectrum is weakly correlated to the total
centre of mass energy of the $\gamma \gamma$ system
because part of the produced particles may go undetected in the
forward and backward regions.
In order to  obtain the W$_{\gamma \gamma}$ distribution
from the \Wvis {} spectrum the method of
\cite{bayes} is used.

After unfolding the events are corrected for  
detector acceptance and efficiency. There is a systematic difference of
$\simeq$ 20\% between the  results obtained with the two Monte Carlo
models, which may be attributed to   a different modelling of the
diffractive interactions which are not easily seen in the detector.

\subsection{Cross sections and systematic errors}

From the number of  events the cross section  \seeh {} is measured. To 
extract the total cross section of two  real photons  the photon flux
\Lgg  {}\cite{Budnev} is calculated using the method described in
\cite{Schuler}  and  the hadronic two-photon cross-section is
extrapolated to zero Q$^2$ using the form factors of  \cite{Sakurai}.
Depending on the form factors used, this calculation may vary by
$\pm$10\% \cite{Schuler}.

\begin{figure}
\centering
  \includegraphics[width=0.75\textwidth,height=0.6\textwidth]{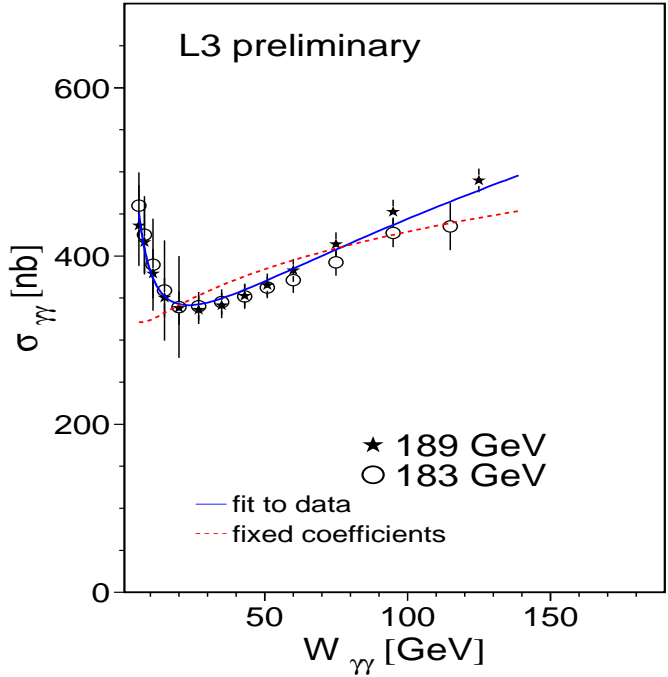}
  \caption{
   The data at \rts = 189 \GeV {} are compared to the data at \rts =
   183 \GeV\ The Regge fits described in the text are superimposed to
   the data corrected with {\sc Phojet}.} 
  \label{eefigb}
\end{figure}

The systematic errors are  evaluated for
each \Wgg {} bin. The main sources  are:  
\begin{list}{$\bullet$}{
\setlength{\topsep}{0pt}
\setlength{\itemsep}{0pt}
\setlength{\parsep}{0pt}
}  

\item uncertainties on the trigger are  5\% at \Wgg = 5 \GeV {} and  
2\% at \Wgg = 145 \GeV. 

\item uncertainties on the energy scale of the calori\-meters excluding 
the clusters measured in  the luminosity calorimeter are $\simeq 1-2
\%$. 

\item  the cut on the number of particles gives a systematic error of
10\% at \Wgg = 5 \GeV {} and $\simeq$ 1\% at \Wgg = 145 \GeV. 

\end{list}   
The mass dependent statistical and systematic errors
are added in quadrature in each \Wgg {} bin. The uncertainty related to
the Monte Carlo model is  not added to the systematic errors, but the
complete analysis is done for {\sc Pythia} as well as for {\sc Phojet}. The
discrepancy on the cross section reflects the difference on selection
efficiency given by the two generators.   The  \ggh {} cross sections
unfolded with {\sc Phojet}
at \rts = 183 \GeV {} and  \rts = 189 \GeV {}   are compared in
Figure~\ref{eefigb}. The data are compatible  within systematic
errors. 

\subsection{Regge parametrisation} 

Total hadronic cross sections show a
characteristic steep decrease in the region of  low centre of mass
energy followed by  a slow rise at high energies. A. Donnachie and P.V.
Landshoff \cite{DL} showed that a parametrisation    of the form 
\begin{equation} \sigma_{\mathrm{tot}}  = A \, s^{\epsilon} \, + \, B
\, s^{-\eta} \label{regge} \end{equation} can account for the energy
behaviour of all  total cross sections, the powers of $s$ being
universal. In a recent compilation of the total
cross section data \cite{PDG98}  a fit of Equation~\ref{regge} for
all hadron total cross sections  gives a result  compatible with the
universal values of  $\epsilon =0.095 \pm 0.002$ and $\eta =0.34 \pm
0.02$. The coefficients $A$ and $B$ are process and $Q^2$ dependent. 

This expression may  also be valid for the two-photon total hadronic
cross section.  The data are fitted to Equation~\ref{regge} with the
parameters  $\epsilon$ and $\eta$  fixed to the world average value.
The coefficients $A$ and $B$ are highly correlated, the correlation
being $\sim$ -0.8. The fit does not represent well  the $\sigma
_{\gamma \gamma}$ energy dependence, we therefore try a fit with $A$,
$B$ and  $\epsilon $ as free parameters. The fits are shown in
Figure~\ref{eefigb} and the results given in Table~\ref{tabletotal}.
The obtained value $\epsilon=0.222\pm0.019$ is  more than a factor two
higher  than the universal value. In PDG 1996 \cite{PDG96} a different
universal fit was given, but with these values the factor two
difference is conserved, showing that  the fitted value of $\epsilon$
is strongly correlated to $\eta$. We try to avoid this by fitting only
the Pomeron exchange for sufficiently high  \Wgg {} values. The 
results, using  different minimum values of \Wgg {} and  different
Monte Carlo generators for unfolding are listed in
Table~\ref{tabletotal}.  We observe also that the value of  $\epsilon$
increases by increasing the lower  mass cutoff, thus indicating a
somewhat variable slope due to the onset of QCD phenomena.

\begin{table}[bpt]
  \begin{center} 
\begin{tabular*}{\textwidth}{c@{\extracolsep{\fill}}cccccc}
  \hline
 fit & \Wgg {} interval& A &  B & $\eta$ fixed&  $\epsilon$ & C.L.  \\                              
  \hline
PDG98 ({\sc Phojet}) &5-145 \GeV& \ 50.$\pm$9.  &1153.$\pm$114. & 0.34& 0.222$\pm$0.019 & 0.995 \\
PDG98 ({\sc Phojet}) &5-145 \GeV& 172.$\pm$3. & \,325.$\pm$65. & 0.34& 0.095 fixed & 0.000034 \\
PDG96 ({\sc Phojet}) &5-145 \GeV& \ \ 90.$\pm$10.  &1519.$\pm$169. & 0.468&
0.168$\pm$0.012 & 0.895 \\
PDG98 ({\sc Pythia}) &5-145 \GeV& \ \ 78.$\pm$10. & \ 753.$\pm$116. & 0.34& 0.206$\pm$0.013 & 0.61 \\ 
\hline 
({\sc Phojet})&13-145 \GeV&\ 230$\pm$10&---&---&0.070$\pm$0.006&$ < 10^{-6}$ \\ 
({\sc Phojet})&31-145 \GeV&150$\pm$8&---&---&0.118$\pm$0.007&0.06 \\ 
({\sc Phojet})&39-145 \GeV&126$\pm$8&---&---&0.136$\pm$0.008&0.45 \\ 
({\sc Phojet})&47-145 \GeV&115$\pm$9&---&---&0.146$\pm$0.010&0.46 \\ 
({\sc Pythia})&13-145 \GeV&185$\pm$7&---&---&0.118$\pm$0.005&$ < 10^{-6}$ \\  
({\sc Pythia})&31-145 \GeV&145$\pm$6&---&---&0.146$\pm$0.006&0.05 \\ 
({\sc Pythia})&39-145 \GeV&135$\pm$7&---&---&0.153$\pm$0.006&0.04 \\ 
({\sc Pythia})&47-145 \GeV&131$\pm$9&---&---&0.156$\pm$0.010&0.14 \\
  \hline
\end{tabular*}
\caption[]{
Fit to the total cross section        of the form \cite{DL} \sgg $= A
\, s^{\epsilon} \, + \, B \, s^{-\eta}$, where $s=W_{\gamma \gamma
}^2$. The statistical and experimental errors and the correlation
matrix between the data points (in total 13 points at \rts=183 \GeV {}
and 13 points at \rts=189 \GeV ) are used in the fit.  The fitted
parameters are strongly correlated. The second set  of fits  evaluates
only the increase of \sgg {} with $s$, i.e. the  Pomeron part of the
fit. {\sc Phojet} and {\sc Pythia} in the first column indicate  the
Monte Carlo used in unfolding the data.
}
  \label{tabletotal}
  \end{center}
\end{table}

\subsection{Models for \gg {} total cross sections}

\begin{figure}
\centering
  \includegraphics[width=0.75\textwidth,height=0.6\textwidth]{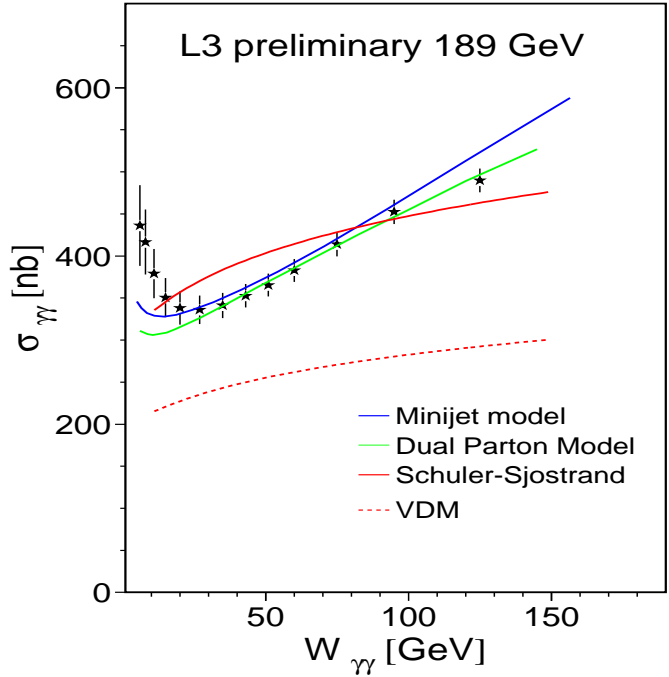}
  \caption{
   Comparison of the data with various models predictions (see text). 
   The data are corrected with {\sc Phojet}. The errors are statistical and systematic added in
quadrature.} 
  \label{eefig}
\end{figure}

Three models describing the total $\gamma\gamma$ cross section are
compared to our data in Figure~\ref{eefig}.

The model of G.A. Schuler and T. Sj\"ostrand \cite{SS} aims at  a
smooth superposition of hadron-like and point-like photon
interactions. Following their discussion, one sees that the
fluctuation of  both photons into  vector mesons is not sufficient to
describe the data. Adding the  point-like splitting of the photon to
q$\bar{\mathrm{q}}$ pairs,  the cross section increases to values
comparable to our data points.

 The cross  sections predicted by R. Engel and J. Ranft \cite{Engel} 
are  in  good agreement  with the
data. In their model   $\gamma$p and pp data are used to fix   the
couplings of the Pomeron and of the Reggeon  to the q$\bar{\mathrm{q}}$
fluctuation of the photon.  The cross sections are then calculated in 
the framework of a Dual Parton Model, with the unitarization
constraint~\cite{Engel}.

The calculation of the mini-jet model of R.M. Godbole and G. Pancheri
\cite{giulia} ascribes to QCD the task of calculating the average
number of semi-hard collisions $n(b,s)$, which is a function of the
impact parameter $b$ and of the centre of mass energy $s$. The
collisions are assumed to be independent of each other at a fixed
value of the impact parameter $b$ and to have a Poisson type
distribution around the average value.  In Figure~\ref{eefig} the
parameters  which fit best the photoproduction data are used
($p_{T\mathrm{min}}=$ 2 \GeV {} and $k_0=$ 0.66 \GeV), together with the GRV
parton density function.

\section{EXCLUSIVE $\rho^0$ PAIR PRODUCTION} 

Since photons can fluctuate directly into neutral Vector Mesons, the
$\gamma\gamma\to\rho^0\rho^0$ cross-section is expected
to remain large  at high
$\gamma\gamma$ centre-of-mass energies~\cite{crossrho} ("elastic"
\gg {} scattering). A measurement
 performed at large $W_{\gamma\gamma}$, above the well known
but still controversial $\rho^0\rho^0$ threshold
enhancement~\cite{twogamrho}, can help to 
understand the behaviour of quasi-elastic and diffractive processes.

The exclusive \eepipi {} events are selected by requiring four tracks,
with a distance of closest  approach  to the  nominal vertex smaller
than 1 mm in the transverse plane, total charge equal to zero and 
missing transverse momentum squared smaller than 0.05 \GeV $^2$.  In the
\rts= 189 \GeV {} sample, we have selected 14000 events. The $4 \pi$
mass spectrum is dominated by the low mass threshold enhancement and
agrees well with  previous measurements \cite{twogamrho}.

\begin{figure}
\centering
\includegraphics[width=0.75\textwidth,height=0.7\textwidth]{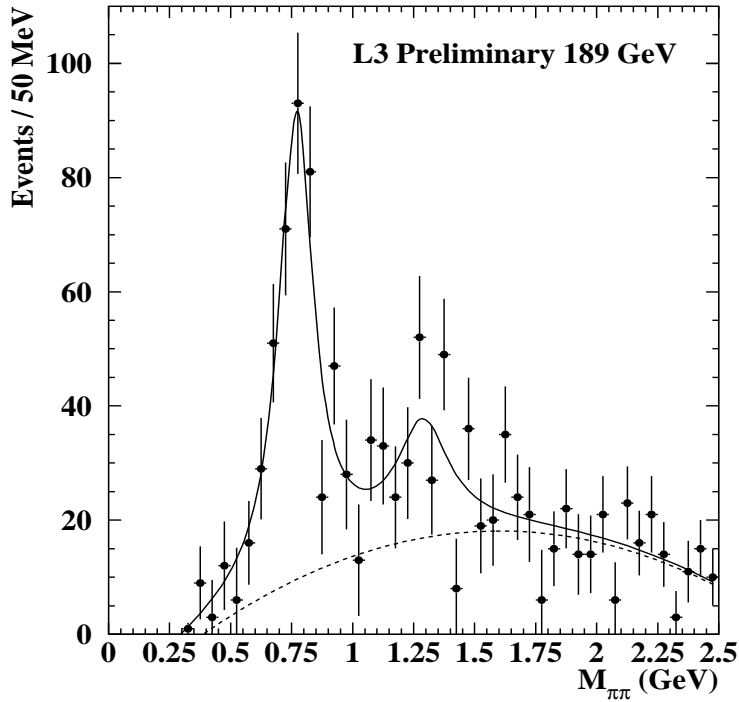}
 \caption{Effective mass of two opposite-charge pions,
 after subtraction of background estimated 
from equal-charge combinations. 
}
  \label{rr2pi}
\end{figure}

\begin{figure}
\centering
 \includegraphics[width=0.75\textwidth,height=0.6\textwidth]{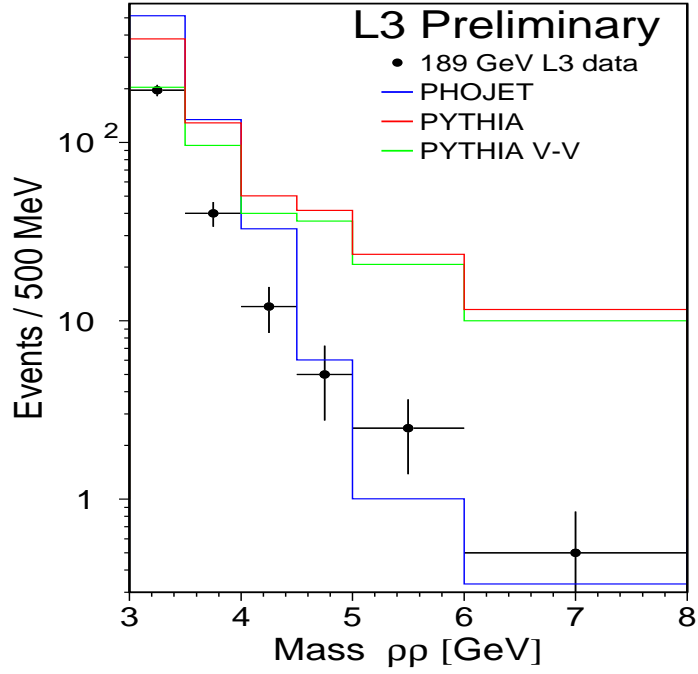}  
  \caption{Number of  $\rho \rho$ events as a function of \Wgg. The data 
are compared to Monte Carlo predictions.}
  \label{rrwgg}
\end{figure}

There are 843 events with a mass \Wgg $ > 3$ \GeV. The four tracks in
each event give rise to four possible $\pi ^+ \pi ^-$ neutral
combinations. The equal charge $\pi ^+ \pi ^+$ and $\pi ^- \pi ^-$
combinations  are used to subtract the combinatorial background. 
Figure~\ref{rr2pi} shows the $\pi ^+ \pi ^-$ mass distribution  after
subtraction  of the equal charge combinations. A peak in the $\rho^0$
mass range is visible, a Breit-Wigner fit gives M$_{\rho}=768 \pm 7 $
\MeV, $\Gamma _{\rho}= 170 \pm 20 $ \MeV. A  second enhancement is
also visible between 1.2 and 1.4 GeV, which corresponds to the
$\mathrm{f}_2$ resonance, the Breit-Wigner fit gives
M$_{\mathrm{f}_2}=1288 \pm 22 $ \MeV, $\Gamma _{\mathrm{f}_2}= 258
\pm 135 $ \MeV. Since the $\mathrm{f}_2$  cannot be produced
diffractively, this indicates the presence of non-diffractive
processes. A $\rho^0\rho^0$ region is defined by requiring both
simultaneous $\pi ^+ \pi ^-$ combinations to have masses below 1.2
\GeV. The \Wgg {} distribution of the  $\rho^0\rho^0$ events  is
shown in Figure~\ref{rrwgg}.  Exclusive  \gg \ra {} VV events were
generated with the {\sc Pythia} Monte Carlo, the dominant channel being \gg
\ra {} $\rho^0\rho^0$.  {\sc Pythia} expects too many events at high
masses, while the mass-dependence of {\sc Phojet} is too steep.

\section{INCLUSIVE $\rho^0$ PRODUCTION}

We study single diffractive dissociation in the inclusive process
$\gamma\gamma\to\rho^0X$ with $\rho^0\to\pi^+\pi^-$ using data
collected  at $\sqrt{s}=183$ GeV, corresponding to ${\cal L}=47.9\
\mathrm{pb}^{-1}$. This process yields a larger number of events
compared to the exclusive $\rho^0\rho^0$ production at the price of
higher background. Higher invariant masses are accessible, but events
can not be fully reconstructed.

We select events with $W_{\mathrm{vis}}> 2$ GeV and $\rho^0$
candidates with $p_T>1$ GeV and $|\cos\vartheta|<0.8$ in the laboratory.
Within this kinematic region the detection efficiency is uniform.

Diffractive events in general contain a rapidity gap that separates the
final state particles originating from the two incoming photons. In
order to separate non-diffractive production, we require the $\rho^0$
candidate to have the largest or smallest rapidity in the event,
excluding its decay products.

Background is subtracted in two steps. First all combinations of two
equal-charge tracks are treated as candidates, subject to the same
cuts, and subtracted from the mass distribution of real candidates with
opposite-charged tracks. The resulting mass distribution is fitted with
a Breit-Wigner curve with fixed parameters $M=770$ MeV and $\Gamma=150$
MeV and a background of the form $e^{A(x-280\
\mathrm{MeV})}(1-e^{B(x-280\ \mathrm{MeV})})$.

\begin{table*}
\centering
\begin{tabular*}{\textwidth}{@{}l@{\extracolsep{\fill}}cccc}
\hline
& $W_{\mathrm{\gamma\gamma}}$ cut
& {$W_\textrm{vis} > 2$ GeV} 
& {$W_\textrm{vis} > 3$ GeV} 
& {$W_\textrm{vis} > 5$ GeV}
\\
\hline
Data &&
202.4 $\pm$ 40.0 & 
\phantom{0}77.1 $\pm$ 23.9  &
 34.3 $\pm$ 6.8
\\
{\sc Pythia} & 3.0 GeV &
123.7 $\pm$ 15.6 &
 108.7$\pm$ 12.5  &
74.3 $\pm$ 8.0
\\
{\sc Pythia} VV & 2.5 GeV &
\phantom{0}98.6 $\pm$ \phantom{0}7.6 &
\phantom{0}82.6  $\pm$ \phantom{0}6.8 &
35.6 $\pm$ 2.3
\\
{\sc Phojet} & 3.0 GeV &
\phantom{0}12.9 $\pm$ 10.9 &
\phantom{00}6.2 $\pm$  \phantom{0}5.9 &
---\\
\hline
\end{tabular*}

\caption[]{Number of inclusive $\gamma\gamma\to\rho^0X$ events after
background subtraction.}
\label{incevents}
\end{table*}

The number of events observed after background subtraction is shown in
Table~\ref{incevents}, together with the numbers expected from three 
Monte Carlo
samples: full {\sc Pythia} $\gamma\gamma\to\textsl{hadrons}$, 
{\sc Pythia} containing only diffractive $\gamma\gamma\to VV$, with
$V=\rho^0,\omega^0,\phi$, and full {\sc Phojet}
$\gamma\gamma\to\textsl{hadrons}$.  The lower value of the generated 
$W_{\gamma\gamma}$ is also given in the table.

\begin{figure}
\centering
\includegraphics[width=0.75\textwidth,height=0.7\textwidth]{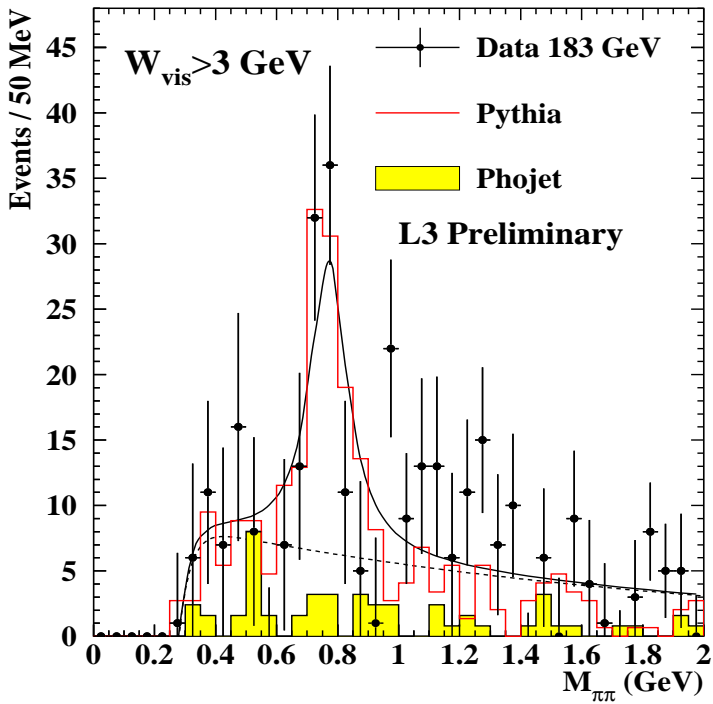}
\caption{Mass distribution of two opposite-charged pions after
subtraction of background estimate from two equal-charged pions. The
data are compared to Monte Carlo predictions and fitted as signal plus
background.}
\label{incmass}
\end{figure}

Figure~\ref{incmass} shows the mass
distribution of the $\rho^0$ candidates for $W_{\mathrm{vis}}>3$ GeV.
The strong $\rho^0$ signal is well reproduced by {\sc Pythia}, while it is
absent in {\sc Phojet}.

To test the diffractive nature of our sample, the decay angle of the
$\rho^0$ candidates is studied. Conservation of the s-channel
helicity implies transverse polarization for diffractive $\rho^0$
production from $Q^2=0$ photons, while non-diffractive $\rho^0$
production is unpolarized.  We define $\vartheta^\ast$ as the angle
between the $\pi^+$ momentum in the $\rho^0$ frame and the $\rho^0$
momentum in the \gg {} rest frame as measured by the visible hadronic
system. This angle is not the true decay angle, but Monte Carlo
simulations show that the $\cos\vartheta^\ast$ resolution is
$\sigma_{\cos\vartheta^\ast}=0.033$. A fit of $\sin^2\vartheta^\ast$,
corresponding to purely transverse polarization, gives $\chi^2=1.82$
(77\% CL), while a fit with a constant, corresponding to unpolarized
decay, gives  $\chi^2=6.18$ and a confidence level of 19\%.

\begin{figure}
\centering
\includegraphics[width=0.65\textwidth,height=0.6\textwidth]{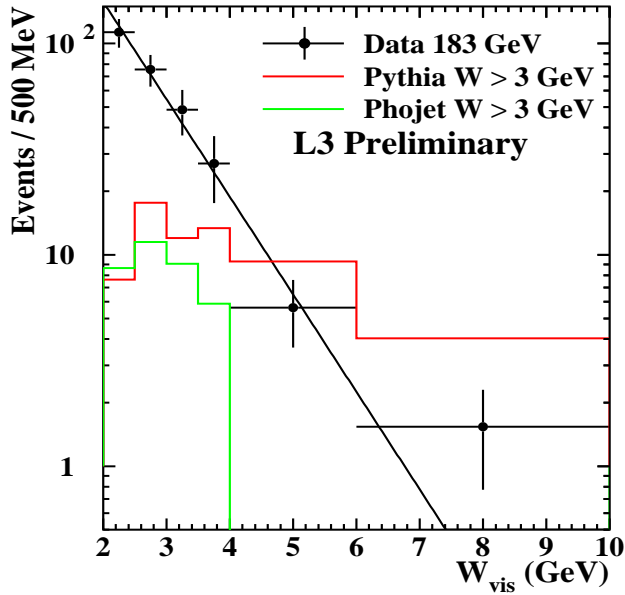}
\caption{Number of  $\gamma\gamma\to\rho^0X$ events as a function of
$W_{\mathrm{vis}}$. The data are compared to the {\sc Pythia} and {\sc Phojet} Monte
Carlo predictions.}
\label{incwvis}
\end{figure}

Figure~\ref{incwvis} shows the number of events observed after
background subtraction as a function of $W_{\mathrm{vis}}$, together
with predictions from {\sc Pythia} and {\sc Phojet}. Note that both Monte Carlo
samples have a generator level cut at $W_{\gamma\gamma}>3$ GeV, so in
the first two bins they are expected to be low. {\sc Phojet} predicts too
few events, while the {\sc Pythia} distribution is too flat. 

The $p_T^2$ distribution  after correction for detector effects using
the sample of vector meson pairs generated by {\sc Pythia} was fitted between
1 GeV$^2$ $\le p_T^2 <$ 3 GeV$^2$ by an exponential function, and the
preliminary slope parameter is $-2.04 \pm 0.11 \GeV ^{-2}$.

\section{CONCLUSIONS}

 The cross section \csee {} for untagged events is measured at LEP,
with the L3 detector, at  \rts\ = 183 and 189 \GeV, in the
interval  5 \GeV $\leq$ \Wgg $\leq$ 145 \GeV. The real photon total
cross section \sggh, derived from the data, increases as a function
of \Wgg, faster than expected from the universal fit of hadron-hadron
total cross-sections.  The observed energy
dependence can be reproduced by QCD models which include hard
scattering of the partons  inside the photon.

 An analysis of \ggrr {} and $\gamma\gamma\to\rho^0X$   events has
been performed. In the region 3 \GeV $\leq$ \Wgg $\leq$ 10 \GeV {} 
both {\sc Pythia} and {\sc Phojet} fail to reproduce the data : the energy
dependence of $\rho^0 \rho^0$ production is too flat in {\sc Pythia} and too
steep in {\sc Phojet}.
 In the  $\gamma\gamma\to\rho^0X$
sample the $ \rho^0$ is transversely polarised as expected  by
s-channel helicity conservation. 
 Almost no $ \rho^0$ is produced by {\sc Phojet} in the
studied kinematical region.

\section*{ACKNOWLEDGEMENTS}

I would like to express my gratitude to  Maria N. Kienzle-Focacci and
Evelyne Delmeire for providing their results and for useful discussions
and encouragement.

This work was partly supported by the Hungarian Scientific  Research
Fund OTKA under contract numbers F-023259 and T-019181. Participation
at this conference was supported by the E\"otv\"os University and the
conference organisers.

\end{document}